\documentclass[prb,aps,twocolumn,showpacs,superscriptaddress]{revtex4}
\usepackage{amsfonts,amsmath,bm}
\newcommand{\rl}{\rangle\!\langle}

\DeclareMathOperator{\tr}{Tr}

\usepackage{graphicx}

\begin{document}

\title{Superradiance and enhanced luminescence from ensembles of a few
  self-assembled quantum dots} 
\author{Wildan Abdussalam}
\affiliation{Institute of Physics, Wroc{\l}aw University of 
Technology, 50-370 Wroc{\l}aw, Poland}
\affiliation{Max-Planck-Institut f\"ur Physik Komplexer Systeme, 
N\"othnitzer Str. 38, 01187 Dresden, Germany}
\author{Pawe{\l} Machnikowski}
 \email{Pawel.Machnikowski@pwr.wroc.pl}  
\affiliation{Institute of Physics, Wroc{\l}aw University of 
Technology, 50-370 Wroc{\l}aw, Poland}

\begin{abstract}
We study theoretically the evolution of photoluminescence (PL) from homogeneous and
inhomogeneous ensembles of a few coupled QDs. We discuss
the relation between signals from a given QD ensemble under strong
and weak excitation (full inversion and linear response regimes): 
A system homogeneous
enough to manifest superradiant emission when strongly
inverted shows a non-exponential decay of the PL signal under
spatially coherent weak excitation. In an inhomogeneous ensemble the PL
decay is always nearly exponential with a qualitatively different
form of the time dependence in the two excitation regimes and with a
higher rate under weak excitation. 
\end{abstract}


\maketitle

\section{Introduction} 
\label{sec:intro}

Optical properties of arrays and ensembles of quantum dots (QDs) continue to
attract attention both from the theoretical
\cite{temnov05,temnov09,averkiev09,yukalov10,auffeves11} and experimental
\cite{scheibner07,mazur09,bogaart10} point of  
view. This interest is certainly motivated to a large extent by the
possible applications, 
in particular in laser structures \cite{ulrich07,boiko12}. However, it
is also driven by purely 
scientific interest in the fundamental properties of these widely
studied systems which still seem to be not completely understood. One
of the currently debated questions is the role of collective
(superradiant) effects in the luminescence of QD ensembles. Signatures
of collective emission were found in the time-resolved
photoluminescence (PL) of planar QD samples \cite{scheibner07} and
QD stacks \cite{mazur09}. In both cases, the coupling between the dots
\cite{mazur09,kasprzak10} seems to be 
essential \cite{kozub12} in order to overcome the detrimental effect
of the ensemble inhomogeneity on the collective dynamics
\cite{sitek07a,sitek09b}. 

In our recent work \cite{kozub12} we were able to propose a model that
reproduced the observed collective enhancement of spontaneous emission
in QD ensembles \cite{scheibner07}. In that study, we assumed weak
excitation of the inhomogeneous QD system and showed, in accordance
with the experimental results, that the collective emission effects
under these conditions are manifested by an increase of the PL decay
rate, while the general form of the time dependence of the PL signal
remains essentially exponential. On the other hand, the usual
superradiance \cite{dicke54,gross82} is observed in strongly excited
(occupation-inverted), 
homogeneous atomic samples, where it is manifested as a delayed, sudden
outburst of radiation from the system \cite{skribanovitz73}, which
therefore shows a markedly non-exponential behavior.

While controlling the degree of initial inversion of a QD ensemble may
be out of question at least at the current stage of development of the
experimental techniques, it seems reasonable to try and extend the
theoretical analysis in order to better relate the QD
``superradiance'' to its atomic prototype. More specifically, it
might be interesting to compare the PL dynamics in inhomogeneous
systems under strong excitation (full occupation inversion) and under
weak excitation (linear response regime), as a function of, e.g., the
size of the ensemble or the degree of inhomogeneity. Based on such
analysis, one 
could be able to predict the behavior of the system in the strong
excitation regime based on the observation of the weak excitation PL
behavior. In particular, it might become clear what kind of behavior a
system should manifest under weak excitation in order to be really
superradiant in the sense of developing the non-monotonic PL response
under (perhaps experimentally unavailable at the moment) strongly inverting
excitation. As an additional benefit, the proposed analysis will allow
us to asses whether the weak excitation assumption made in the
previous work \cite{kozub12} did not suppress the PL decay, thus
forcing us to introduce the short-range coupling that might turn out
to be spurious if stronger excitation is assumed.

Thus, in this paper, we study the collective spontaneous emission from small
ensembles of coupled QDs, comparing the time-resolved photoluminescence signal
in the cases of strong excitation (full inversion) and weak resonant
excitation. We show that the buildup of the superradiant emission peak
in a strongly inverted sufficiently homogeneous QD ensemble correlates
with the clearly non-exponential PL decay under weak
excitation of the same ensemble. On the other hand, for more
inhomogeneous ensembles (including the currently realistic ones), in
both excitation regimes the decay of the PL signal may be
indistinguishable from exponential. In the latter case, rather
surprisingly, the collective enhancement of the decay rate is stronger
under weak excitation.

The paper is organized as follows. In Sec.~\ref{sec:model}, we
describe the model of a small ensemble of QDs and the method of
simulation. Next, in 
Sec.~\ref{sec:results},  we present and discuss the simulation results for the
photon emission under different excitation conditions. Finally, Sec.~\ref{sec:concl} 
concludes the paper.

\section{Model} 
\label{sec:model}

We consider a planar, single-layer ensemble of a few (up to eight)
self-assembled QDs randomly and uniformly placed in the sample plane
($xy$ plane in the model). The model of the ensemble closely follows
that of our previous work \cite{kozub12}: The positions of the dots are denoted
by $\bm{r}_{\alpha}$, where $\alpha$ numbers the dots. 
We introduce the restriction that the
center-to-center distance between the QDs can not be lower than
10~nm (roughly the QD diameter).
Each QD 
is modeled as a point-wise two-level system (empty  
dot and one exciton) with the fundamental transition energy
$E_{\alpha}=\overline{E}+\epsilon_{\alpha}$, where $\overline{E}$ is
the average transition energy in the ensemble and $\epsilon_{\alpha}$
represent the energy inhomogeneity of the ensemble, described by a
Gaussian distribution with zero mean and standard deviation $\sigma$. 
Following our previous findings \cite{kozub12}, we assume 
the dots to be coupled by an interaction $V_{\alpha\beta}$ which is composed
of long-range (LR) dipole interaction (dispersion force) and a
short-range (SR) coupling (exponentially decaying with the distance),
\begin{displaymath}
V_{\alpha\beta}=V^{(\mathrm{sr})}_{\alpha\beta}+V^{(\mathrm{lr})}_{\alpha\beta}.
\end{displaymath}
The long-range dipole coupling is described by 
\cite{stephen64,lehmberg70a,varfolomeev71,kozub12}
\begin{displaymath}
V^{(\mathrm{lr})}_{\alpha\beta}
=-\hbar\Gamma_{0}G(k_{0}r_{\alpha\beta}),\quad \alpha\neq\beta,
\end{displaymath}
and $V_{\alpha\alpha}=0$,
where $\bm{r}_{\alpha\beta}=\bm{r}_{\alpha}-\bm{r}_{\beta}$,
$\Gamma_{0}=
|d_{0}|^{2}k_{0}^{3}/(3\pi\varepsilon_{0}\varepsilon_{\mathrm{r}})$
is the spontaneous emission (radiative recombination) rate for a
single dot, $d_{0}$ is the magnitude of the interband dipole moment
(assumed identical for all the dots),
$\varepsilon_{0}$ is the vacuum permittivity,
$\varepsilon_{\mathrm{r}}$ is the relative dielectric constant of the
semiconductor, $k_{0}=n\overline{E}/(\hbar c)$,
$c$ is the speed of light,
$n=\sqrt{\varepsilon_{\mathrm{r}}}$ is the refractive index of the
semiconductor, 
and, for a heavy-hole transition in a planar ensemble,
\begin{displaymath}
G(x)  = -\frac{3}{8}\left(
\frac{\cos x}{x}+ \frac{\sin x}{x^{2}}+\frac{\cos x}{x^{3}}\right).
\end{displaymath}
For the SR coupling, which plays a much more important role
\cite{kozub12}, only the overall magnitude and finite range are
important, hence we model it by the simple exponential dependence
\begin{displaymath}
V^{(\mathrm{sr})}_{\alpha\beta}
=V_{0}e^{-r_{\alpha\beta}/r_{0}}.
\end{displaymath}

The equation of evolution of the density matrix is then given
by \cite{lehmberg70a,kozub12} 
\begin{equation}
\label{evol}
\dot{\rho}=-\frac{i}{\hbar}[H_{0},\rho]+
\mathcal{L}[\rho].
\end{equation}
Here the first term accounts for the unitary evolution of the ensemble
of coupled QDs with the Hamiltonian
\begin{displaymath}
H_{0} = \sum_{\alpha =1}^{N} \epsilon_\alpha \sigma_\alpha^\dagger
\sigma_\alpha 
+ \sum_{\alpha, \beta = 1}^N 
V_{\alpha \beta} \sigma_{\alpha}^{\dagger} \sigma_{\beta},
\end{displaymath}
where we introduce the transition
operators for the dots: the ``exciton annihilation'' operators
$\sigma_{\alpha}$ which annihilate an exciton in the dot $\alpha$,  
and the ``exciton creation'' operators $\sigma_{\alpha}^{\dagger}$
which creates an exciton in the dot $\alpha$
(the exciton number operator for the dot $\alpha$ is then
$\hat{n}_{\alpha}=\sigma_{\alpha}^{\dag}\sigma_{\alpha}$). 
In the standard basis, the operators $\sigma^{\dag}_{\alpha},\sigma_{\alpha}$ correspond to
the raising and lowering operators and can be represented
by Pauli matrices on a given two-level (pseudospin) system,
$\sigma_{\alpha}=(|0\rl 1|)_{\alpha} = (1/2)(\sigma_{x}-i\sigma_{y})_{\alpha}$.
The second term describes the dissipation, that is, the collective
spontaneous emission process due to the coupling between the quantum
emitters (QDs) and their radiative environment (vacuum). This is
modeled in terms of the dissipator
\begin{displaymath}
\mathcal{L}[\rho] =
\sum_{\alpha,\beta=1}^{N}\Gamma_{\alpha\beta}\left[ 
\sigma_{\alpha}\rho\sigma_{\beta}^{\dag}
-\frac{1}{2}\left\{ \sigma_{\beta}^{\dag}\sigma_{\alpha},\rho \right\}
 \right].
\end{displaymath}
Here $\Gamma_{\alpha\alpha}=\Gamma_{0},\quad
\Gamma_{\alpha\beta}=\Gamma_{\beta\alpha}=\Gamma_{0}F(k_{0}r_{\alpha\beta})$,
with 
\begin{eqnarray*}
F(x) & = & \frac{3}{4}\left(
\frac{\sin x}{x}-\frac{\cos x}{x^{2}}+\frac{\sin x}{x^{3}}
 \right),
\end{eqnarray*}
and $\{\ldots,\ldots\}$ denotes the anti-commutator. 

Note that, although our equations lead to a numerically exact solution
within the proposed model, the density matrix formalism restricts the
available information to quantum-mechanical averages, hence some
aspects of the quantum dynamics, like, e.g., the
field fluctuations that trigger the superradiance on the very short
time scales \cite{yukalov10}, although present in the underlying
microscopic physics, cannot be explicitly
accounted for in our approach.

The simulations are performed by randomly placing a given number of
QDs with a fixed surface density $\nu$ in the $xy$ plane, choosing
their fundamental transition energies from the Gaussian distribution,
and then directly numerically solving
Eq.~\eqref{evol}. 
Depending on the excitation conditions, a broad variety of initial
states can be thought of, with subsequent dynamics depending on the
amount of inversion as well as on the degree of spatial coherence
induced by the excitation. Here, we restrict our discussion to the two
extreme cases: a fully inverted or weakly excited
initial state. The former is a product state (without spatial coherence
or correlation between the dots) characterized by the highest possible
degree of excitation (exciton number).
In terms of our notation, this fully inverted initial state corresponding to strong
excitation conditions is
\begin{displaymath}
|\Psi_{0}^{\mathrm{(FI)}}\rangle = 
\prod_{\alpha=1}^{N}\sigma_{\alpha}^{\dag}|\mathrm{vac}\rangle,
\end{displaymath}
where $|\mathrm{vac}\rangle$ is the ``vacuum'' state, that is,
the crystal ground state with filled valence band states and
empty conduction band states (no excitons in the QDs). In the case of
a weakly excited ensemble, the essential feature is the spatial
coherence between the QDs, which forms naturally when the whole
ensemble is coherently and resonantly illuminated but seems to
appear also under quasi-resonant excitation conditions
\cite{scheibner07}. The equations of motion for exciton
occupations (that govern the PL signal) decouple from the evolution of
\textit{interband} coherences and, when admitting at most one exciton
in the system, the total signal is simply proportional to the initial
average occupation. Hence, as our initial state reflecting the weak
excitation conditions we formally take the coherently delocalized single-exciton state
\begin{displaymath}
|\Psi_{0}^{\mathrm{(WE)}}\rangle = 
\frac{1}{\sqrt{N}}\sum_{\alpha=1}^{N}\sigma_{\alpha}^{\dag}
|\mathrm{vac}\rangle.
\end{displaymath}
This state is an equal superposition of states, each of which has a single
emitter (QD) inverted, hence it contains one exciton and will lead to
emission of a single photon
(thus effectively normalizing the signal to unit initial occupation).

In our discussion, we focus on the time evolution of the
total PL intensity, that is the photon emission rate or, equivalently,
the exciton number decay rate. From Eq.~\eqref{evol}, this is given in
therms of the density matrix by
\begin{displaymath}
I = -\frac{d}{dt}\sum_{\alpha}
\langle\sigma^{\dag}_{\alpha}\sigma_{\alpha}\rangle
= -\sum_{\alpha}
\tr \left( \mathcal{L}[\rho] \sigma^{\dag}_{\alpha}\sigma_{\alpha}\right) 
\end{displaymath}

In our simulations, we  use the parameters for a CdSe/ZnSe QD system:
$\Gamma_{0}=2.56$~ns$^{-1}$, $n=2.6$, the average transition energy of
the QD ensemble $\overline{E}=2.59$~eV and the QD surface density 
$\nu=10^{11}$ /cm$^{-2}$. For the
tunnel coupling we choose the amplitude $V_{0}=5$~meV and the range
$r_{0}=15$~nm, which are the values used in the previous work
\cite{kozub12} to reproduce the experimental results \cite{scheibner07}.

\section{Results}
\label{sec:results}

In this section we present the results of our simulations of the
time-resolved PL from ensembles of a few QDs under different
excitation conditions. In each case, we performed 100 simulations for
ensembles with different spatial and spectral distributions of the QDs
and, subsequently, averaged the results.

\begin{figure}[tb]
\begin{center}
\includegraphics[width=85mm]{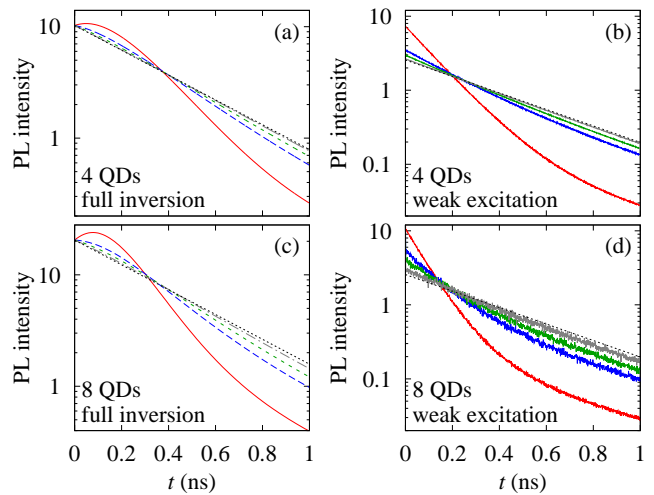}
\end{center}
\caption{\label{fig:sigma}The PL signal under strong (a,c) and weak (b,d)
  excitation for an ensemble of 4 (a,b) and 8 (c,d) QDs, as a function
of the ensemble inhomogeneity: $\sigma=0,5,10,30$~meV for solid red,
dashed blue, dashed green, and dash-dotted grey lines,
respectively. The black dotted lines show the exponential decay of PL
from a single dot. All the curves are averaged over 100 QD ensembles.}
\end{figure}

In Fig.~\ref{fig:sigma}, we show the time dependence of the PL signal
from systems of 4 and 8 QDs with a varying degree of spectral
inhomogeneity. Under strong excitation, when the initial state is
fully inverted (left panels), non-monotonic development of the PL
signal is visible in the case of perfectly homogeneous ensembles (red solid
lines), corresponding to the superradiant peak that would develop much
more clearly in larger ensembles. A weak non-monotonicity is still
visible for a weakly inhomogeneous ensemble ($\sigma=5$~meV, blue
dashed line), while for the more inhomogeneous ensembles the PL
decay is monotonic. Although the qualitative form of the time
dependence of the PL signal becomes hardly distinguishable from
exponential, the decay is noticeably faster than that corresponding to
a single QD (shown with a black dotted line). Apart from a more
pronounced maximum and a faster decay in the case of 8 QDs, there is
no qualitative difference between the two ensemble sizes. Under weak
excitation (linear response regime, right panels in
Fig.~\ref{fig:sigma}), the PL decay is always monotonic. However, in a
homogeneous or sufficiently weakly inhomogeneous system (red and blue
lines) the PL decay is non-exponential. In a larger system this
non-exponential behavior extends to larger values of inhomogeneity
[green line, corresponding to $\sigma=10$~meV, in
Fig.~\ref{fig:sigma}(d)] but eventually, for strongly inhomogeneous
systems the decay also becomes exponential. Let us note at this point
that in the experimentally studied ensemble \cite{scheibner07}, one
had $\sigma\approx18$~meV.

Comparison of the simulated PL dynamics in a fully inverted system
[Fig.~\ref{fig:sigma}(a,c)] with that under weak excitation
[Fig.~\ref{fig:sigma}(b,d)] leads to the first main conclusion of our
analysis.
A given system can emit in a different way depending on the initial
excitation. In practice, the control of excitation conditions (initial
state) may be limited. E.g., it may be hard to induce full inversion
of all the QDs in the ensemble. Our results allow one to infer the
dynamics of the system under full inversion based on the PL decay from
the same system under weak excitation: If the system,
when fully inverted, is able to show peaked,
superradiant emission, then it manifests its superradiant properties already
in the linear response (weak excitation) regime by a non-exponential
decay of the PL signal. Conversely, a system in which, when excited
weakly, the PL signal decays exponentially will show a monotonic
decay, close to exponential, under strong inversion. In fact,
especially for a system of 8 QDs, the weak excitation
dynamics is non-exponential already for $\sigma=10$~meV, while the
decay under full inversion in such an ensemble is monotonic and rather
close to exponential apart from the initial phase of a few tens of
picoseconds. Thus, the conditions for developing actual superradiance (in
particular the spectral homogeneity of the system) are more
strict than those allowing deviations from exponential decay in the
linear response regime. It may also be interesting to note that the PL
decay in inhomogeneous systems in the two excitation regimes, even
though close to exponential 
in both cases, is still qualitatively different: the time dependence of the
PL signal, when plotted in the logarithmic scale, is concave for
strong inversion and convex in the linear response regime, at least
for sufficiently short times (on the order of the PL life time).

\begin{figure}[tb]
\begin{center}
\includegraphics[width=85mm]{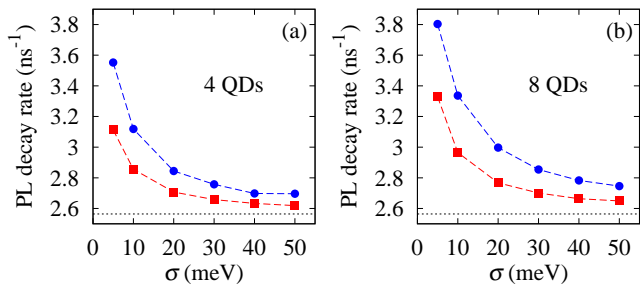}
\end{center}
\caption{\label{fig:times}PL decay rates extracted from exponential
  fitting to the PL decay curves for 4 (a) and 8 (b) QDs. Blue circles
  and red squares correspond to weak and strong excitation regimes,
  respectively. Dashed lines are added to guide the eye. Black dotted lines
  show the decay rate for a single QD.}
\end{figure}

In view of the fact that the PL decay in an inhomogeneous system is
close to exponential and can be indistinguishable from the latter
based on actual experimental data it seems reasonable to extract the
apparent decay rates from the PL evolution by fitting the PL
curves with an exponential dependence. The result for the two ensemble
sizes discussed above and for a series of values of $\sigma$ is shown
in Fig.~\ref{fig:times}, where the blue circles and red squares
correspond to weak and strong excitation, respectively. This result is
the second main conclusion of this work: the PL 
signal under weak excitation decays faster than after fully inverting
the system. This means that the spatial coherence generated by the
global state preparation is higher than that achievable spontaneously
by the inhomogeneous system in the evolution of the inverted state
(contrary to the standard, highly symmetric case of non-interacting,
identical atoms \cite{nussenzweig73}, where the system evolves via the
subspace of maximally spatially coherent states). 

\begin{figure}[tb]
\begin{center}
\includegraphics[width=85mm]{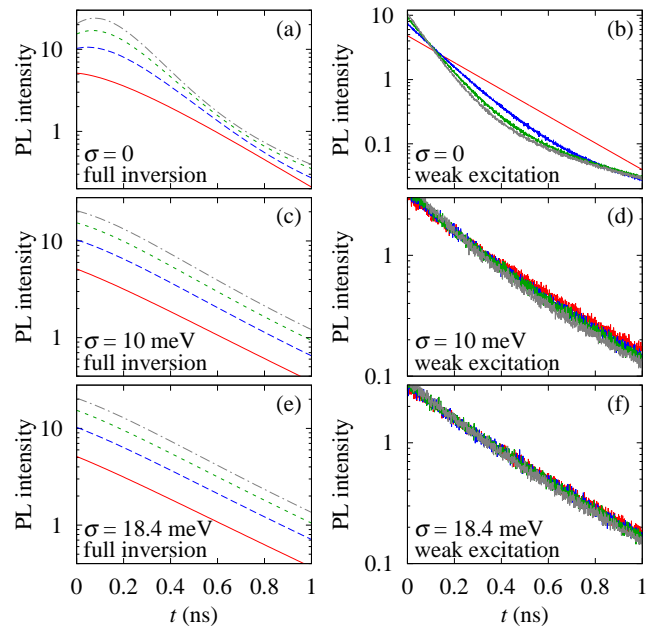}
\end{center}
\caption{\label{fig:number}The time-resolved PL signal as a function
  of the QD ensemble size for three values of the ensemble
  inhomogeneity as shown and for the fully inverted (a,c,e) and weakly
  excited (b,d,f) initial state. Red solid: 2 QDs, blue dashed: 4 QDs, green
  dashed: 6 QDs, grey dash-dotted: 8 QDs.}
\end{figure}
 
For the sake of completeness, let us conclude our discussion with a
brief analysis of how the PL signal evolves with growing ensemble
size. The pertinent simulation results are presented in
Fig. \ref{fig:number}. In a strongly inverted homogeneous ensemble
[Fig.~\ref{fig:number}(a)], non-monotonicity develops already for 3
or 4 QDs, in accordance with earlier findings for regular QD arrays
\cite{sitek07a}. This is again reflected by non-exponential decay
under weak excitation [Fig.~\ref{fig:number}(b)]. For a realistic
degree of inhomogeneity [Fig.~\ref{fig:number}(c-f)], the decay is
only weakly non-exponential. For strong excitation, the PL intensities
mostly differ by their magnitude (proportional to the number of the
QDs) with only a small variation of the shape (flattening of the
curve) at short times, which becomes stronger in larger ensembles. In
a weakly excited inhomogeneous ensemble, the PL decay curves almost
overlap and the difference of the decay rate becomes unnoticeable for
$\sigma=18.4$~meV, which roughly corresponds to the ensemble studied
in the experiment \cite{scheibner07} (nonetheless, more careful
quantitative analysis shows that the rates do increase with the
ensemble size \cite{kozub12}).
   
\section{Conclusions}
\label{sec:concl}

We have modeled the evolution of the PL signal from homogeneous and
inhomogeneous ensembles of a few (up to 8) coupled QDs. We focused on
the comparison between the PL response under strong excitation (fully
inverted initial state) and weak excitation (linear response). 

We have
shown that the signals from a given QD ensemble in the two regimes,
although obviously different, are correlated: A system homogeneous
enough to manifest non-monotonic, superradiant emission when strongly
inverted shows a non-exponential decay of the PL signal under
spatially coherent weak excitation. In a more inhomogeneous system the
PL decay under weak excitation is close to exponential and so is the
time-resolved PL signal under full occupation inversion. The QD
samples in which collective emission was found experimentally
\cite{scheibner07} belong to the latter class.

While the PL decay converges to the simple exponential form as the
inhomogeneity grows, it retains a different character in the two
excitation regimes, showing concave and convex behavior (in the
logarithmic scale) for strong and weak excitation, respectively.

Quantitatively, when fitting the nearly exponential PL decay with a
strictly exponential dependence, the decay under weak excitation
appears faster than in the fully inverted case. Hence, simulations
performed for weakly excited systems (which are much less demanding
computationally) yield an upper bound on the apparent decay rates for
a given system under any excitation intensity.

\textbf{Acknowledgment:} This work was supported in parts by the
Polish National Science Centre (Grant No. DEC-2011/01/B/ST3/02415) 
and by the Foundation for
Polish Science under the TEAM programme, co-financed by the European
Regional Development Fund. 


\end{document}